\documentclass{article}
\usepackage[a4paper, total={6in, 8in}]{geometry}
\usepackage[utf8]{inputenc}
\usepackage[numbers]{natbib}
\usepackage{amsmath}
\usepackage{url}
\usepackage{graphicx}
\usepackage[]{hyperref}
\usepackage{breakurl}
\usepackage{cleveref}
\usepackage{orcidlink}
\usepackage{tikz}
\usepackage{color}
\usetikzlibrary{positioning,patterns,decorations.pathreplacing,calc}
\usepackage[framemethod=TikZ]{mdframed}
\usepackage{newfloat}

\DeclareFloatingEnvironment[fileext=frm,placement={!b},name=Frame]{myfloat}
\newenvironment{infobox}[2][]{%
    \begin{myfloat}[#1]
    \begin{mdframed}[%
        frametitle={Information: #2},
        skipabove=\baselineskip plus 2pt minus 1pt,
        skipbelow=\baselineskip plus 2pt minus 1pt,
        frametitleaboveskip= 7pt,
        frametitlebelowskip= 7pt,
        linewidth=1pt,
        linecolor=black,
        frametitlerule=false,
        frametitlebackgroundcolor=blue!10,
        backgroundcolor=blue!10,
        roundcorner=7pt,
    ]%
}{\end{mdframed}\end{myfloat}}

\newcommand*{\doi}[1]{\href{https://doi.org/#1}{https://doi.org/#1}}
\crefformat{footnote}{#2\footnotemark[#1]#3}
\title{Cool URIs for FAIR Knowledge Graphs%
}
\author{Andreas Thalhammer\textsuperscript{\orcidlinkf{0000-0002-0991-5771}}}
\date{Version 1 - June 18, 2024}%
\begin{document}

\maketitle

\section{TL;DR}
\label{sec:tldr}
\begin{enumerate}
    \item Own the URI if you want to make own claims about the resource; avoid creating claims about URIs that belong to someone else.
    \item Use \texttt{https://}; avoid \texttt{http://}
    \item Make your URIs dereferencable.
    \begin{enumerate}
         \item Offer the identified resources in one or more machine-readable format(s) via content negotiation; avoid ``HTML only''.
         \item Use \textit{slash URIs} in combination with 303 redirects and metadata; avoid hash URIs.
         \item Decommission URIs with \textit{301 moved permanently} or \textit{410 gone}; avoid \textit{200 ok} and \textit{404 not found} in these scenarios.
         \item Use a central redirect service; avoid multiple custom subdomains.
        \item Avoid authentication and, where indispensable, make it easy and ensure that dereferencing continues after the authentication challenge.
    \end{enumerate}
    \item Use opaque short and stable paths; avoid organizational, technology, taxonomy, or ontology-related cues in the folder(s)/context and accession identifiers.
    \begin{enumerate}
        \item Use unreserved and unambiguous ASCII characters in paths; avoid characters from the broader UTF-8 spectrum and ambiguity.
        \item Use only lowercase in the path before the accession identifier; avoid changes between lowercase and uppercase.
        \item The accession identifier marks the end of the path; avoid trailing slashes and technology-related extensions such as ``.html'' or ``.ttl''.
    \end{enumerate}
    \item Use opaque numeric or alphanumeric accession identifiers; avoid mnemonic accession identifiers.
\end{enumerate}
Please find more information in the individual points in the respective subsections of Section \ref{sec:guide}. Prefixes used in this document are available in Appendix~\ref{apx:prefix}.\\
\\

\noindent\textbf{Disclaimer}: Most topics covered in this work haven been subject to controversy. What you find here is just the opinion of the author.

\section{Introduction}
\begin{figure}
\begin{centering}
\begin{tikzpicture}
  \node[draw] (scheme){\texttt{https://}};
  \node[draw] (subdomain) [right = of scheme] {\texttt{purl.}};
  \node[draw] (domain) [right = of subdomain]{\texttt{example.com}};
  \node[draw] (path) [right = of domain]{\texttt{/a9/}};
  \node[draw] (accession) [right = of path]{\texttt{e42}};
  \node[align=center] (draw) (path1) [below of = path]{Folder(s) or \\context};
  \node (scheme1) [below of = scheme] {Scheme};
  \node (subdomain1) [below of = subdomain] {Subdomain};
  \node (domain1) [below of = domain] {Domain};
  \node[align=center] (accession1) [below of = accession] {Accession\\identifier\footnotemark};
  \node (path1) at ($(path)!0.4!(accession)$) {};
  \node (path2) [above of = path1] {Path};
  \draw [
    thick,
    decoration={
        brace,
        raise=0.5cm
    },
    decorate
] (path.west) -- (accession.east) ;
\end{tikzpicture}
\caption{Example of a ``Cool URI for a FAIR Knowledge Graph''.}
\label{fig:example}
\end{centering}
\end{figure}
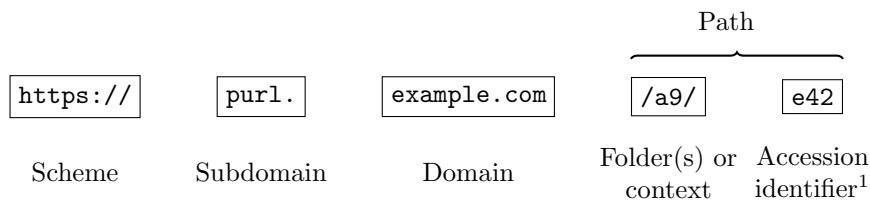
\footnotetext{We derive this term from library science/museum inventories/archaeology where it is frequently called ``accession number''. It is usually an alphanumeric identifier that is locally controlled by (part of) an organization. McMurry et al. and others call it ``local identifier/local ID'' \cite{McMurry117812}.}
This guide is for everyone who seeks advice for creating stable, secure, and persistent Uniform Resource Identifiers (URIs) in order to publish their data in accordance to the FAIR principles \cite{fair}. The use case does not matter. It could range from publishing the results of a small research project to a large knowledge graph at a big corporation. The FAIR principles apply equally and this is why it is important to put extra thought into the URI selection process. The title aims to extend the tradition of ``Cool URIs don't change'' \cite{Berners-Lee1998} and ``Cool URIs for the Semantic Web'' \cite{Sauermann}. Much has changed since the publication of these works and we would like to revisit some of the principles. Many still hold today, some had to be reworked, and we could also identify new ones.

Knowledge Graph enthusiasts will quickly realize that our example in Figure~\ref{fig:example} strongly resembles the URI style that Wikidata offers and, indeed, there is \texttt{http://www.wikidata.org/entity/Q42}. We argue that Wikidata, back in 2012/2013, got most parts perfectly right. As a point of criticism, considering the evolution of the web in the last decade, the chosen scheme for Wikidata should have been \texttt{https://} (more on this topic in Section \ref{subsec:https}).\footnote{\label{ft:denny}\url{https://phabricator.wikimedia.org/T258590\#6333873}} Also, more marginally, the accession identifier could also have been lower case, just like the rest of the path, and consecutive counting is discouraged in some scenarios (see Section \ref{subsubsec:opaque}).
Now, only getting one-and-a-half things ``wrong" is, in fact, very good. The motivation for this guide comes from a number of mistakes that we made in the past couple of years and the list is far more extensive:
\begin{itemize}
    \item We identified core-business objects with the URIs of someone else.
    \item We used hash URIs and tried to redirect them later.
    \item We offered HTML only.
    \item We used mnemonic accession identifiers.
    \item We used authentication that broke dereferenceability.
    \item We maintained a custom subdomain with a redirect after the subdomain of ``the system'' has changed.
    \item And yes, we also used \texttt{http://} instead of \texttt{https://}.
\end{itemize}
Most of this was either ``unFAIR'', inefficient, ineffective, or just not very sustainable. However, we are all allowed to make mistakes so we can learn and grow. We came up with the quick guide in Section \ref{sec:tldr} and provide more detail in the following sections. We hope that the reader of this guide can avoid some of our mistakes.

\section{Guide}
\label{sec:guide}
In the tradition of ``Cool URIs'' \cite{Berners-Lee1998,Sauermann} we provide guidance on technical design in addition to syntax as the former influences the latter. 

\subsection{Ownership}
\label{subsec:own}
Many resources that deserve a URI already exist and the ``R'' in FAIR stands for ``reuse''(-able). We could decide to utilize already existing URIs such as \texttt{uniprot:P37231} or \texttt{wikidata:Q154842} for our applications. However, there are scenarios where we want to create our own URIs for exactly those resources. We  then can (and should) link them to existing, publicly available, ones. There are two scenarios that indicate when we should be ready to create our own URIs:

\begin{infobox}{DOIs, CURIEs and LSIDs}
In this work we are, in fact, more precisely talking about ``Cool URLs" rather than ``Cool URIs" (URLs are a subclass of URIs). If a URL contains a hostname separated with a dot from a top-level domain (instead of a local name or an IP address) we are call it a ``fully qualified domain name'' and directly rely on the  Domain Name System (DNS) system to resolve it. DNS has always played a key role for ``Cool URIs'' \cite{Berners-Lee1998,Sauermann}. It could be argued that identification and domain ownership should be separate things (``who owns SEPTIN2 anyways?'') but this is where we leave the ground of ``Cool URIs'' and enter the space of the Handle system and DOIs \cite{handle}, CURIE \cite{curie} registries/resolvers (e.g., \url{https://bioregistry.io/}), or LSIDs \cite{lsid}. What these systems have in common is that they don't rely on the DNS system for registration/resolution and therefore bring their own systems. Common issues that these systems face include coordinating allocated sub-spaces as well as semantic meaning among different providers. In the course of time they were either made compatible to URLs (i.e., DOIs) or designed to be compatible (i.e., CURIEs and LSIDs). As such, or in the form of URNs, they are also usable in an RDF context.
\end{infobox}

\begin{enumerate}
    \item When the identified resource is part of \textbf{core business.}
    \item When we want to make own claims about the resources.\footnote{\label{ft:usinguris}That is, not just using the URI in a referential way (such as a pointer or citation). ``To make a claim about a resource" typically includes using the URI in the subject position of the created triples.}
\end{enumerate}
Commonly the first implies the second. Sometimes there are peripheral resources that need to be described, e.g., to specify provenance or further context where only the second point holds. The main reasons for having own URIs for core business objects include protection of proprietary information, gained flexibility, and development speed. Especially the last point may seem counterintutive but there are teams that start rushing with ``let's be quick and use UniProt'' and get stuck with ``we sometimes also need to create own entities because UniProt does not cover them (yet); how will we reconcile our data when UniProt finally has a URI for that protein?'' By reusing someone else's identifier for our ``thing we want to describe'' we are tying our life cycle to an external lifecycle---and lose flexibility. Which bring us to ``core business'': It includes everything that an organization/application primarily focuses on. For example, if an application focuses on searching/browsing art works, then core business objects may include artists, art works, series, genres, techniques, materials, galleries, museums, etc. There are other resources such as locations or weather conditions that could be sourced from existing knowledge graphs in this scenario. When we decide to coin our own URIs, we should link to existing resources with one of \texttt{owl:sameAs}, \texttt{schema:sameAs}, \texttt{skos:exactMatch}, \texttt{skos:closeMatch}, \texttt{skos:related}, or \texttt{rdfs:seeAlso}---depending on the degree of similarity \cite{halpin_identity}. At this point, it is important to be as clear as possible about the semantics of the new resource from the very beginning by adding clear definitions and/or ``anchoring'' it with \texttt{owl:sameAs}. 

There is also the notion of authoritative/non-authoritative publication \cite{hogan2010weaving} of triples. This means, for example, that we should not publish triples that describe/re-define\cref{ft:usinguris} the Uniprot resource \texttt{uniprot:P37231} on our server via the document URI \texttt{https://purl.example.com/a9/doc/e42.ttl} because the authoritative\footnote{The document to which \texttt{uniprot:P37231} points when dereferenced. See Section \ref{subsec:dereferenceability} for more information on dereferencability.} document for that resource is located at the UniProt server in a document named \texttt{P37231.ttl} (see Section \ref{subsec:dereferenceability} for more information on resources vs. documents). Put in a simple way: it does not make sense to publish a document that is all about \texttt{uniprot:P37231} on a server that is not related to the UniProt project in any way. To our data this does more harm than good: a Web crawler will not be able to retrieve our information by looking up the identifier. Similarly, considering the FAIR principles  \cite{fair}, the \textit{``A1. (meta)data are retrievable by their identifier using a standardized communications protocol''} principle implies that the ``data we publish" is retrievable when we call an entity's URI via HTTP(s). This is certainly not the case when we use someone else's URI to identify our entity. Finally, it is definitely better to create a new identifier and then sort out the relation to existing ones (see above for property suggestions) than re-using an existing identifier in a wrong or unintended way, realizing about it afterwards, and then start a dissection/cleanup exercise.

Within larger organizations it is normal that multiple URIs exist for the same or similar concepts. This is often the case when different systems focus on different aspects of a resource. For instance an employee's salary statement is stored in a different system than the one that tracks their office location. In these instances it is often worthwhile to ask the following questions:
\begin{enumerate}
    \item Should the additional information that one system brings about a resource type feed back to the main\footnote{``Main'' in the sense of ``where the resources are created and their basic information is kept up-to-date for the whole organization``.} system of record for these resources? %
    
    \item If the previous question is answered with ``no'': are there cross-links (e.g., \texttt{skos:closeMatch}) so that the respective records can be discovered from either system? %
\end{enumerate}
We can put the above ideas into the context of ``high cohesion / low coupling'' which should be familiar concepts to most software and enterprise architects. The decision of ``URI re-use versus create" also depends on whether we stay within a certain ``Bounded Context'' - another relevant concept we borrow from software architecture.\footnote{``Bounded Context" is a core concept of Domain-Driven Design \cite{ddd}.}

Further reading:
\begin{itemize}
    \item The FAIR Cookbook highlights the context set by a source for an entity: ``The need for each database to mint their own identifier is a result of each taking a different perspective on the concept. [...] These differences in perspective are driven by the goal of the databse [sic] in terms of the data that they store.''\footnote{\url{https://web.archive.org/web/20230715093743/https://faircookbook.elixir-europe.org/content/recipes/interoperability/identifier-mapping.html}} This matches the above-mentioned concept of Bounded Context borrowed from Domain-Driven Design \cite{ddd}.
    \item Bess Schrader on re-using your URI: ``While there’s nothing technically stopping someone else from re-using this URI in a knowledge graph they create, they would be unable to have that URI resolve to anything other than my URI page, which would alert them that this URI is already in use. It’s somewhat equivalent to someone else signing up for an account with your email address – they can do it, but they’ll be unable to receive messages or reset their password, so it doesn’t make much sense.''\footnote{\url{https://web.archive.org/web/20221208172422/https://enterprise-knowledge.com/resolving-uniform-resource-identifiers-uris/}}
    \item The Cabinet Office of the British government provides guidance for ``Designing URI sets for the UK public sector''. It states: ``When considering the domain to root a URI set in: the publisher will require content control of the sub-domain that it ultimately resolves to''~\cite{ukgov}.
    \item The BBC implemented an interesting approach in their collaboration with MusicBrainz where they re-used the accession identifiers but contextualized them with their own domain and folder structure.\footnote{For example: \url{https://web.archive.org/web/20180829012720/https://www.bbc.co.uk/music/artists/20244d07-534f-4eff-b4d4-930878889970}}. This could be perceived as a good compromise but it can bring problems if we want to ``extend offline" without creating redundancy or conflicts with the future evolution of the source.
    \item Figure 3 of \cite{McMurry117812} highlights the trade-off between ``Create new ID vs reuse existing''. The authors' general advise in their Lesson 1 is ``to  only create your own identifiers for new knowledge''~\cite{McMurry117812}. 
    
\end{itemize}
\subsection{https://}
\label{subsec:https}
First things first: Are \begin{itemize}
    \item \texttt{http://example.com} and
    \item  \texttt{https://example.com} 
\end{itemize}technically different URIs? Yes, they are.\footnote{See Section 4.2.2 of RFC 9110 \cite{HTTPSemantics} where it says: ``Resources made available via the `https' scheme have no shared identity with the `http' scheme. They are distinct origins with separate namespaces.''} Can I introduce HTTP URIs and then fix it with \texttt{owl:sameAs} to link to the respective HTTPS counterpart?\footnote{As stated before, a real counterpart does not exist: there is no rule that these two URLs need should represent the same. We are relying on goodwill.} Yes but no: Consumers may still try to resolve the plain HTTP URIs (which is a major security issue \cite{Halpin:PrivOn2017}). Welcome to the section where a tiny ``s" makes a big difference.

At some point this topic was subject to strong debates but many practitioners have already moved to the ``https'' scheme and others are either in the process or would like to migrate. Here are some facts that will help us to understand the context and the various arguments.
\begin{itemize}
    \item Port 80 (plain HTTP) communication is not considered secure as information is sent in plain text across the internet. Already a single initial request via plain HTTP can be subject to snooping and man-in-the-middle attacks.
    \item Port 443 (HTTPS) communication is the current standard for secure interaction between web clients and web servers based on Transport Layer Security (TLS). Modern browsers use the ``s" in the scheme of the URI as a cue for whether a secure connection should be established upfront and also use the provided certificate to verify authenticity via a chain of trust\footnote{\url{https://en.wikipedia.org/wiki/Chain_of_trust}}. In addition, HSTS and according preload lists are also a way to make things more secure (see next point).
    \item HSTS \cite{HSTS} is a policy that enables strict use of TLS after the first request/response interaction via HTTPS. It works by setting the ``Strict-Transport-Security" HTTP response header field. When browsers encounter this field all future requests to the web server will automatically use TLS even if links are specified with the ``http'' scheme. If web servers implement this correctly the traffic becomes mostly secure. One limitation is that the first request/response interaction can still be in plain HTTP. Modern browsers mitigate this by utilizing HSTS preload lists\footnote{\url{https://hstspreload.org/}} to enable security also for the first request.
\end{itemize}
Based on these facts, there are a number of arguments for using plain \texttt{http://} URIs:
\begin{enumerate}
    \item The ``s" in the URI should not make the difference: TLS should be the default technology for HTTP communication and security should never have been pushed up to the application level. Changing existing URIs to \texttt{https://} will have more downsides than upsides.\footnote{\url{https://www.w3.org/DesignIssues/Security-NotTheS.html}}
    \item Traditionally, vocabularies like RDF and RDFS don't use the \texttt{https} scheme. It could confuse users to see multiple prefix definitions: some with \texttt{https://} and  others with \texttt{http://} (e.g., in a SPARQL query or Turtle file). In any case, URIs are just identifiers, not links.\footnote{\label{ft:phabricator}\url{https://phabricator.wikimedia.org/T153563}}
    \item If your domain offers HSTS and is in the HSTS preload list of most modern browsers, no user will actually send a plain HTTP request.\cref{ft:phabricator} Let the infrastructure solve it.\footnote{\url{https://www.w3.org/blog/2016/05/https-and-the-semantic-weblinked-data/}}
\end{enumerate}
Now we analyze these arguments one by one:
\begin{enumerate}
    \item It is probably too late to push TLS down the layer stack to where it belongs. Things like ``HSTS preload lists'' would not exist if there was an easy solution. In addition, in this work, we are not talking about legacy URIs but want to provide guidance on how to coin new, fresh ones.
    \item  The idea of the Semantic Web, Linked Data, and also the FAIR guidelines is that identifiers are meant to be dereferencable: users and machines can and should use them in a ``follow your nose'' pattern (see Section \ref{subsec:dereferenceability} for more information on dereferenceability). It appears random to tell our users ``Hey, here is a URI where you can find more information, but don't use it, it's just an identifier (\textit{psst, it's insecure})". Regarding consistency: Semantic Web standards evolved over a period of 20 years; one prefix using \texttt{https://} and the other \texttt{http://} does probably not qualify for the hall of fame of ``biggest inconsistencies''.
    \item We could coin \texttt{http://} URIs and try to make them secure by using HSTS including preload lists, changing application code where possible, and other mitigation measures. We treat the symptoms; the actual problem would only be solved if we would coin \texttt{https://} URIs. Even worse, most mitigation strategies are primarily targeting human users but do not prevent a machine with \texttt{curl}\footnote{curl does not and probably will not support the type of preload list that other modern browsers currently use: \url{https://daniel.haxx.se/blog/2020/11/03/hsts-your-curl/}} to slowly resolve 20, 50, or 1000s of plain HTTP URIs that were discovered, for instance, in a Turtle file via (semantic) web crawling (see below for examples) or returned by a SPARQL endpoint. This is the opposite to what Semantic Web, Linked Data, and FAIR commonly stand for, i.e., treating machines as first-class citizens on the web: TLS is de-facto not in place in such scenarios and every single request poses multiple threats to the user. Finally, when we offer \texttt{http://} URIs with HSTS support, there is a likelihood that we may actually get it (partially) wrong.\footnote{\url{https://github.com/perma-id/w3id.org/issues/1942}}
\end{enumerate}
When we talk about this aspect it is important to note that, in fact, we address a topic that often lies in the policy rather than the technical space: 301 redirects from plain HTTP to HTTPS are enabled in a vast majority all cases.\footnote{This is due to the general recommendations on the implementation of HSTS although the HTTP to HTTPS redirect is not strictly part of the HSTS specification \cite{HSTS}.} So it is more about which URIs we communicate to our users. We could even be ambiguous on this topic and accept both ways like, for example, \texttt{schema.org}.\footnote{\url{https://schema.org/docs/faq.html\#19}}
However, this aspect becomes important when \texttt{https://} URIs are preferred, HSTS is enabled, and 301 redirects from plain HTTP to HTTPS are implemented. This, for example, holds for all URIs served via \url{https://w3id.org}. Although the main idea of this service is ``... to provide a secure, permanent URL re-direction service for Web applications''\footnote{https://w3id.org/} it does not prevent users to deliberately or accidentally use \texttt{http://w3id.org/} in their prefix definitions (and potentially publish thousands of insecure URIs on the web).\footnote{For example the prefix definitions of \url{https://w3id.org/fraunhofer/lighthouse-projects/evolopro/cirp.ttl}, \url{https://w3id.org/quality/qskos}, or \url{https://w3id.org/glosis/model/codelists}. Checked on \today{}, the git commit history of \texttt{w3id} will show changes/fixes that happened afterwards (they are/would be appreciated).} In case of a mistake, it may go unnoticed because the URIs of the defined linked data resources do resolve (just with one more 301 redirect at the beginning). This is strongly related to Hyrum's Law\footnote{\label{hyrum}\url{https://web.archive.org/web/20240601204008/https://www.hyrumslaw.com/}} and should make maintainers of such a redirect service aware that enabling 301 redirects from plain HTTP to HTTPS basically means: users will use and depend on \texttt{http://} as a scheme for their URIs (although otherwise intended in the case of \url{https://w3id.org}). A viable measure to prevent or detect incorrect use faster could be to disable 301 redirects from HTTP to HTTPS.\footnote{Note: HSTS can still be enabled in such scenario as the redirect is not part of this policy~\cite{HSTS}.} In this case, URIs with \texttt{http://} as a scheme could still be created and shared but, at least, they would not resolve. This should be considered when hosting a central redirect service for your own organization (see Section \ref{subsubsec:central_id}).

We conclude with an overview of multiple arguments for coining URIs with the ``https'' scheme instead of ``http'':
\begin{itemize}
    \item Harry Haplin investigated the topic in ``Semantic insecurity: Security and the Semantic Web"~\cite{Halpin:PrivOn2017}.
    \item Jana Hentschke provides summary of a workshop titled ``Cool URIs might be insecure if they don't change" (including nice material and observation collections).\footnote{\url{https://web.archive.org/web/20230715104411/https://wiki.dnb.de/display/DINIAGKIM/HTTP+vs.+HTTPS+in+resource+identification}}
    \item Dan Brickley on the topic of migrating \texttt{schema.org} to HTTPS: ``Switching the content of the JSON-LD context to generate https: triples is a very special situation. [...] If we do go there, I think it's the kind of change we ought to publicize at least a year in advance, with significant supporting documentation.''\footnote{\url{https://github.com/schemaorg/schemaorg/issues/2853\#issuecomment-810100754}}
    \item Denny Vrande{\v{c}}i{\'{c}} on the topic for Wikidata/Commons: ``In my opinion, I would suggest to use the chance and since we are building a new resource here, to use https from the start. It was a mistake that we didn't do that for Wikidata, which was due to the fact that Wikipedia only started offering https in 2011, and switched to it only in 2015. We are not in that situation today, and can make the right decision. I would love for Wikidata to switch, but that's a different task."\cref{ft:denny}
\end{itemize}

\subsection{Dereferenceability}
\label{subsec:dereferenceability}

\begin{infobox}{httpRange-14 and ``real-world resources''}
The ``correct'' way of distinguishing  between real-world resources and the documents describing them was likely the biggest controversy in the history of the Semantic Web. At some point it received the name ``httpRange-14''. It has occupied hundreds of researchers and software architects and has been re-opened and re-discussed at various points over the past 20 years.\footnote{\url{https://web.archive.org/web/20230715104823/https://www.w3.org/wiki/HttpRange14Webography}} ``On the Semantic Web, URIs identify not just Web documents, but also real-world objects like people and cars, and even abstract ideas and non-existing things like a mythical unicorn. We call these real-world objects or things.'' \cite{Sauermann} When we talk about real-world resources we basically mean anything that is not a web document.\footnote{The distinction between real-world resources and the documents describing them is often also called ``non-information resources vs. information resources'' \cite{httpRange-14}.} %
\end{infobox}
When we coin new URIs we need to consider two main aspects:
\begin{enumerate}
    \item How do we clearly separate but still relate the identified resource from/to the (web) document(s) describing it?
    \item How do we reconcile between the human-readable document and the machine-readable (web) document(s) that describe the resource?
\end{enumerate}
The first point was the subject of a decade-long discussion (and the resolution is still rather sub-optimal; see below for more information) while the solution for the second point was basically built into the early versions\footnote{\url{https://www.w3.org/Protocols/HTTP/1.0/spec.html}, \texttt{Accept} and \texttt{Content-Type} header fields.} of the HTTP protocol.
We will now outline what they respectively mean for the example URI in stated in Figure~\ref{fig:example}, \texttt{https://purl.example.com/a9/e42}:
\begin{enumerate}
    \item Assume the URI describes a real-world entity, e.g., the person ``Douglas Adams''. In this case, we need a separate URI to identify the document that contains information about the real-world entity. Why? Because we may want to attach metadata about the document itself, e.g., add the creator of the document, its last change, etc. If this separation \textbf{does not} happen, we may end up with the following two triples:\\
\texttt{ex:e42 dbo:birthDate "1952-03-11"\^{}\^{}xsd:date .}\\
\texttt{ex:e42 dcterms:modified "2020-12-25"\^{}\^{}xsd:date .\\}
    It would only be guessable from the used properties which triple describes the person ``Douglas Adams'' and which triple describes the document (that itself contains a description of ``Douglas Adams''). Therefore, we need a different URI for identifying the document itself; it is also a resource\footnote{An information resource, to be precise.}. Even if we do not want to add any metadata about the document at the point of creation we may want to do so in the future.
    \item In many cases we want to support different human and machine-readable formats for the description of our resource; or at least reserve the possibility to add them at a later point in time. Therefore we need to think about a way to distinguish them in a consistent way. In particular, we would like to be in a position where the human browser, asking for \texttt{text/html}, automatically receives \texttt{https://purl.example.com/a9/doc/e42.html} while a machine agent is able to receive the \texttt{text/turtle} representation located at the alternative address\\ \texttt{https://purl.example.com/a9/doc/e42.ttl}.
\end{enumerate}
Both of the above points get interrelated when the requester (human or machine) initially asks to get\footnote{Literally via ``HTTP GET'' request.} \texttt{https://purl.example.com/a9/e42}. At first, we need to signal to the requester that we are dealing with a real-world resource that can not be sent through the ether. But then, as a second step, we may want to be able to offer a web document that describes the real-world resource in the format specified in the (original) \texttt{Accept} HTTP request header.

In addition to the above issues we also cover the following related topics in this section: URI decommissioning, hosting a central identifier/redirect service, and authentication.

\subsubsection{Real-world Resources}
\label{subsubsec:303}
Let's start with the first issue: How do we separate real-world resources from information resources (documents) on the web? 
In June 2005, the W3C Technical Architecture Group (TAG) presented a feasible solution for the above issues that would not rely on the use of the fragment part of URIs (see below for fragement/hash URIs) to identify real-world resources.\footnote{\url{https://lists.w3.org/Archives/Public/www-tag/2005Jun/0039.html}}
Sauerman et al. later summarized this in their ``Cool URIs for the Semantic Web" guide \cite{Sauermann} which is considered the main reference to resolve this type of problem. The solution basically consists of the use of 303 redirects in combination with HTTP content negotiation. The ``303 See Other'' HTTP status code indicates that the resource identified by the requested URI is, in fact, a real-world thing that can not be sent through the ether. As 3xx HTTP codes specify redirects the response will include a \texttt{Location} header field where the location of a (or another) web document is provided. In the case of a 303 redirect, that web document is meant to include a description of the originally requested real-world resource. For more information on the details of this mechanism the reader is referred to sections 4.2 and 4.3 of \cite{Sauermann}. 

Unfortunately, despite being an elegant solution to a complex problem, the 303 approach has never really scored well in terms of usability. This is mostly due to the fact that the publisher needs to control part of the internals of a web server. In addition, there are parts that are not compatible with the original idea of 3xx redirects. Wikidata (and many others) serve as an example for this: If, on the page \url{https://www.wikidata.org/wiki/Q116812461} we right-click on the left second-last link named ``Concept URI" and copy the URL we get \texttt{wikidata:Q116812461}. This is the URI that identifies the real-world resource. However, if we resolve this URI with \texttt{curl} and ask for \texttt{text/turtle} we get the following chain of redirects:

\begin{footnotesize}
\begin{verbatim}
GET http://www.wikidata.org/entity/Q116812461
-->
HTTP/1.1 301 TLS Redirect
Location: https://www.wikidata.org/entity/Q116812461
-->
HTTP/2 303
Location: https://www.wikidata.org/wiki/Special:EntityData/Q116812461
-->
HTTP/2 303
Location: https://www.wikidata.org/wiki/Special:EntityData/Q116812461.ttl
-->
HTTP/2 200
\end{verbatim}
\end{footnotesize}
So this first 301 redirect is telling us: ``301 Moved Permanently''. This literally means: ``don't use that URL any longer: we have a new one now at \textgreater Location\textless ; please use that one''. This is great---see Section \ref{subsec:https} and Section \ref{subsubsec:decommission}. However, by definition, as a machine user this should also mean that the document we will retrieve about the real-world resource (after the 303 redirect) should describe the real-world resource with the URI to which the 301 redirect points to (the server told us we should not use the original URI any longer). However, this is not the case: the entity URI has not changed in the document. The document uses \texttt{wikidata:Q116812461}. For a human user that just has right-clicked on ``Concept URI" and copied the link (containing http without an ``s'') this may be obvious but a machine has no options to dissect this. In other scenarios, there may be multiple 301 redirects and it ends in a guessing game to find out which is the one used in the document---if at all: in case we enter the redirect chain from in-between (which, actually, would be considered good practice after we have seen ``301 Moved Permanently"), say starting with \url{https://www.wikidata.org/entity/Q116812461}, there is no way to know what the original URL was in the first place. In Wikidata this is addressed by adding metadata that comes in the form of triples:

\begin{footnotesize}
\begin{verbatim}
<https://www.wikidata.org/wiki/Special:EntityData/Q116812461> a schema:Dataset ;
	schema:about wikidata:Q116812461
	...
\end{verbatim}
\end{footnotesize}
To this point, the only way to figure ``what this document is actually about'' is to use the URL returned by the first 303 redirect in the subject position, fix the property to \texttt{schema:about}, and the object will contain the URL of the real-world resource that the document describes. It is probably not much of a surprise that this mechanism is specific for Wikidata and will be different for other publishers.

The above example shows that we are still far from an optimal solution regarding \texttt{httpRange-14}. Having identified that issue, Ian Davis proposed Tucan publishing\footnote{\label{ft:tucan}\url{https://web.archive.org/web/20220121005220/http://blog.iandavis.com/2010/11/is-303-really-necessary}}  as a mitigation in 2010. The main idea is to drop 303 redirects altogether and just use a standard property---\texttt{wdrs:describedby}---in the document to denote
\begin{itemize}
    \item what is the URL of the real-world resource and
    \item what is the document describing it.
\end{itemize}
To some extent, this is exactly how Wikidata also mitigates the above-mentioned redirect-chain issue: they use the \texttt{schema:about} property (see above)---but in addition to a 303 redirect, not as a replacement (replacement is suggested in Tucan publishing). However, we believe that both proposed properties, \texttt{schema:about} (Wikidata, see above) as well as \texttt{wdrs:describedby}\cref{ft:tucan}, are overloaded and not specifically designed for this purpose. In the course of writing this document it came clear to us that it may be beneficial to come up with a new property to break down the often complex redirect chain\footnote{It still should include a 303 somewhere.} as single triple in the returned document as metadata. We call the property \url{https://w3id.org/303} and its primary idea is to augment the (often complex) redirect chain. We acknowledge that the existence of any single property does not solve the problem at hand and it probably requires some community discussion and eventually consensus of ``what could be a viable solution here''. We will be happy to update this guide with a subsequent version when a better solution is found. So let us conclude this part by quoting Tim Berners-Lee: ``I am happy to look at improvements in the current HTTPRange 14 architecture as I know that 303 is a pain.  But I don't want to break the web.''\footnote{\url{https://lists.w3.org/Archives/Public/public-lod/2010Nov/0554.html}} 

When we read ``Cool URIs for the Semantic Web'' \cite{Sauermann} there is also Section~4.1 that offers an alternative way to 303 redirects: hash URIs. The nice thing about this identifier scheme is that it ``naturally'' relates entities and the documents describing them. Entities are identified by using the fragment part of the URI while the part before the fragment identifies the document describing them:
\begin{itemize}
    \item \texttt{https://purl.example.com/a9\#e42} $\leftarrow$ real-world resource.
    \item \texttt{https://purl.example.com/a9} $\leftarrow$ document describing the  real-world resource.
\end{itemize}
The most problematic implication that comes with this scheme is that identifiers, once grouped before the fragment, can not be separated afterwards. This is due to the specification of URI fragments: they are stripped off before the HTTP request is sent to the server. Therefore, should ever come up a good reason {not to} serve descriptions of the two entities \texttt{https://purl.example.com/a9\#e42} and \texttt{https://purl.example.com/a9\#e43} together in the same document, we would not be able to make it work: what arrives at the server is the request for \texttt{https://purl.example.com/a9}. Therefore, no current or future redirect engine can treat \texttt{e42} and \texttt{e43} separately. This removes a lot of flexibility from the very beginning and we recommend to avoid creating hash URIs (even for vocabularies/ontologies).

\subsubsection{Content Negotiation}
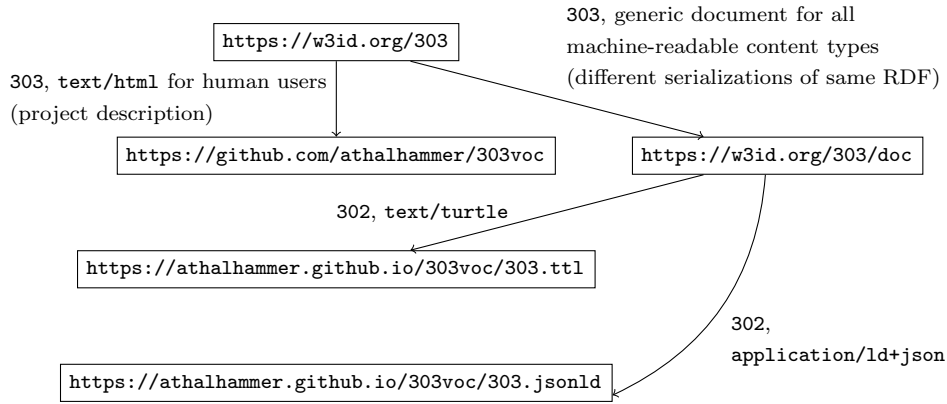
\begin{figure}
\begin{centering}
\begin{tikzpicture}
    \node[draw] (slashdir) {\footnotesize{\texttt{https://w3id.org/303}}};
  \node[draw] (html) [below = of slashdir]{\footnotesize{\texttt{https://github.com/athalhammer/303voc}}};
  \node[draw] (gendoc) [right = of html]{\footnotesize{\texttt{https://w3id.org/303/doc}}};
  \node[draw] (ttl) [below = of html]{\footnotesize{\texttt{https://athalhammer.github.io/303voc/303.ttl}}};
  \node[draw] (jsonld) [below = of ttl] {\footnotesize{\texttt{https://athalhammer.github.io/303voc/303.jsonld}}};
   \draw[->]  (slashdir) edge node[left,align=left] {\footnotesize{\texttt{303}, \texttt{text/html} for human users}\\ \footnotesize{(project description)}} (html);
   \draw[->]  (slashdir) edge node[above right,align=left] {\footnotesize{\texttt{303}, generic document for all}\\\footnotesize{machine-readable content types}\\\footnotesize{(different serializations of same RDF)}} (gendoc);
   \draw[->]  (gendoc) edge  node[left=5mm]{\footnotesize{\texttt{302}, \texttt{text/turtle}}} (ttl);
   \draw[->]  (gendoc) edge[bend left]  node[below right,align=left]{\footnotesize{\texttt{302},} \\ \footnotesize{\texttt{application/ld+json}}} (jsonld);
\end{tikzpicture}
\caption{Example: Implementation of 303 redirects with \texttt{w3id.org} for the \url{https://w3id.org/303} property.}
\label{fig:redirect}
\end{centering}
\end{figure}

In the above example of Wikidata's redirect chain we can see that eventually a Turtle document is returned. This is expected as the client asked for the \texttt{text/turtle} MIME type in its "Accept" request header. The mechanism with which the server handles this request is called ``content negotiation". In our case, where we already have to deal with ``real-world resource'' vs ``document describing it'', content negotiation adds another dimension. Fortunately, it can be resolved quite easily in two different ways: either with or without the introduction of a generic document (or even a combination of the two; see the example in Figure \ref{fig:redirect}). Sauermann et al. \cite{Sauermann} recommend introducing an generic document if the returned data is the same but the format is different. If there are differences in the data itself they recommend to skip the generic document and perform the 303 redirect and content negotiation in a single step. In the above Wikidata redirect chain the generic document is identified with 

\begin{footnotesize}
\begin{verbatim}
https://www.wikidata.org/wiki/Special:EntityData/Q116812461
\end{verbatim}
\end{footnotesize}
This URI also occurs in the finally returned Turtle document where it is of RDF type \texttt{schema:Dataset}. The second 303 redirect in the above Wikidata example was probably introduced for technical reasons. The original idea of this form of content negotiation was to directly return \texttt{200 OK} when the generic document was requested along with a ``Content-Location" header for future reference. Unfortunately, as mentioned in Section \ref{subsubsec:303}, there are often many server-side limitations for the publisher that make implementing all the recommendations difficult. Figure \ref{fig:redirect}\footnote{The full code can be found in the repositories \url{https://github.com/perma-id/w3id.org/tree/master/303} and \url{https://github.com/athalhammer/303voc}.} provides an example for such a trade-off: As \texttt{w3id.org} does not support acting as a proxy\footnote{\url{https://github.com/perma-id/w3id.org/pull/136\#issuecomment-144165179}} we added 302 redirects to the respective data representations instead of directly returning \texttt{200 OK} along with the \texttt{Content-Location} header. We regard the 302 redirect as a good compromise  as we keep only a single 303 redirect in the chain and do not potentially mislead (machine) users by using a \texttt{301 Moved Permanently} redirect (see Section \ref{subsubsec:decommission} for more information). In addition, as mentioned before, the example also shows how the two ways of content negotiation---with or without generic document---can be combined in a straight-forward manner as. In this case, the human representation leads to description of the project while the different machine-readable formats are 1:1 conversions between the documents containing the RDF descriptions of \url{https://w3id.org/303} in different serializations. For more information on content negotiation in combination with slash (and also hash) URIs the reader is referred to \cite{Sauermann}.

In summary, we recommend the following:
\begin{enumerate}
    \item Separate the URIs of the real-world entity and the documents describing them, e.g.:\begin{itemize}
        \item \texttt{https://purl.example.com/a9/e42} $\leftarrow$ real-world resource.
        \item ( \texttt{https://purl.example.com/a9/doc/e42} $\leftarrow$ generic document. )
        \item  \texttt{https://purl.example.com/a9/doc/e42.html} $\leftarrow$ HTML representation of the document describing the real-world resource.
        \item \texttt{https://purl.example.com/a9/doc/e42.ttl} $\leftarrow$ Turtle representation of the document describing the real-world resource.
    \end{itemize}
    When we talk about ``Cool URIs'' we do actually only mean the URIs of the real-world resource; so there is some freedom in how exactly the URIs of the documents describing them look like and also whether there should be a generic document or not. %
    \item Use content negotiation to redirect or directly deliver the requested format (if you can). Make sure to deliver the correct media type in the \texttt{Content-Type} field in the HTTP header when a document is returned.
    \item The redirect chain should include only one \texttt{303, See Other} redirect in case a real-world object was originally requested and no such redirect if a document was requested directly.
    \item When the document describing the real-world resource is delivered in RDF it should use the real-world resource's URI in the contained triples to describe it, e.g.\\
    \texttt{https://purl.example.com/a9/e42\\
    dbo:birthDate "1952-03-11"\^{}\^{}xsd:date .\\}
    In order to add metadata about the document itself it should use the document's URI accordingly, e.g., \texttt{https://purl.example.com/a9/doc/e42\\dcterms:modified "2020-12-25"\^{}\^{}xsd:date .}
    \item Use metadata inside the returned document to state clearly which URI is used to describe the real-world resource and which URI is used to describe the information resource (both resources are described inside the finally returned document). This could be done with a single triple that uses the \url{https://w3id.org/303} property, e.g.
 
    \noindent
    \begin{footnotesize}
\begin{verbatim}
https://purl.example.com/a9/e42 
    https://w3id.org/303 https://purl.example.com/a9/doc/e42 .
\end{verbatim}
\end{footnotesize}
\end{enumerate}

\subsubsection{URI Life Cycle}
\label{subsubsec:decommission}
Like software, data, planets, and stars, URIs and the resources they identify also have a life cycle. Apart from existence versus non-existence, there are three main events that can happen to a resource:
\begin{enumerate}
    \item They are not relevant any longer (no successor).
    \item Their understanding has changed:
    \begin{enumerate}
        \item They morph to another resource (straight-forward direct successor).\footnote{This scenario also applies when the URI needs to be changed for technical/practical reasons.}
        \item They split into a set of other resources (it's complicated, no straight-forward successor).
    \end{enumerate}
\end{enumerate}
For these cases we provide similar recommendations as implemented with Glottocodes \cite{glotto} and provided by \cite{McMurry117812}. When a resource is not relevant any longer (Event 1) we recommend to use a tombstone document delivered with the HTTP response status code \texttt{410, Gone}. The document should include a human readable explanation for the deprecation. The same method should be used when there is no straight-forward successor (Event 2b) and a human is needed to sort out the issue (e.g., ``a partner company has split into two parts - to which part did our contract that we have with them move?''). Although the original concept has a relation to the successors it depends on the context to decide how to proceed (a machine often can not disentangle this).

It is different when a resource has morphed into a straight-forward successor concept (Event~2a). This happens, for example, when two companies merge (both of the original URIs will have a direct successor). The scenario also applies if the URI of a resource needs to change for technical or practical reasons. The server should offer a \texttt{301, Moved Permanently} redirect. A consumer should interpret it in the way: ``kindly use the following URI for future pointers to the real-world resource for any reference; we will do the same''. 

Unfortunately, especially for real-world resources, \texttt{301} redirects do not always comply with that statement. For example, the first redirect of the Wikidata redirect chain has the function to switch to TLS-secured communication while the returned RDF document two redirects later uses the \texttt{http://} URI to describe the concept. We would expect  \texttt{https://} because, literally, it did say ``Moved Permanently''. How ``permanent'' is this move if two redirects later we receive a document that has the old URI in it? Therefore, as mentioned in Section \ref{subsubsec:303}, we recommend to be explicit about the URI of the real-world resource in the returned (RDF) data itself, e.g., by using the \texttt{https://w3id.org/303} property. 

In the case of a straight-forward successor (indicated by HTTP response status code 301), additionally, we recommend to explicitly specify that ``has moved" fact in the returned RDF as well as in the human-readable representation. Suitable properties are \texttt{dcterms:replaces} and \texttt{dcterms:isReplacedBy}; we recommend to include both directions in the returned document.

Finally, there are cases when a real-world resource does not exist (yet) and has never existed before. In that case we recommend the use of the \texttt{404 Not Found} HTTP response status code.

\subsubsection{Hosting a Central Redirect Service}
\label{subsubsec:central_id}
What fits into place with the previous points on 303 redirects and content negotiation is the following: if we need redirects in any case, why don't we directly centralize them? Redirect services such as \url{https://w3id.org/} can help to provide long-term support for URIs even if the domain name of our institution changes. If, for proprietary reasons, an institution can not rely on a public redirect service we recommend to have a central place where all identifiers are bundled, e.g., \texttt{https://purl.example.com}. For example, a frequently occurring pattern is that URIs are coined with the introduction of new system that is available at, e.g., \texttt{https://sys.example.com}. Later, the system is migrated to a new domain or even replaced, to/by e.g., \texttt{https://sys2.example.com}. All old URIs, e.g., \texttt{https://sys.example.com/e42}, now need to redirect to the new location. The old subdomain as well as the according DNS entry, certificates, web server then also need to be maintained in order to have stable, dereferenceable URIs (all of this with the assumption that the DNS record can not be changed). If this scenario happens multiple times within the same organization this can lead to two results:
\begin{enumerate}
    \item Stable URIs are not guaranteed (usually this topic is not on top of the priority list during system migrations).
    \item Guaranteeing stable identifiers for all systems across multiple migrations amounts to significant cost.
\end{enumerate}
Both scenarios pose a challenge. The root of the issue lies in the tight coupling of identifiers and systems that host the descriptions of the real-world resources. This is somewhat logical as these systems usually provide a content management component that enables create, modify, and delete functionalities for the resource descriptions.  A central redirect service can decouple the identifiers of the real-world resources from the systems that host their descriptions. 
We can reserve a path, e.g. \texttt{https://purl.example.com/a9/.*}, which can then 303-redirect to the \texttt{https://sys.example.com/doc/\$1} target. The migration of a system to a new subdomain can then be performed by simply adjusting the redirect: the original identifiers remain stable (the accession identifiers usually don't change during such migrations) The additional level of indirection also helps to prevent typical mistakes related to dereferenceability, for example, not separating the identified real-world entity from the document containing a description of it (see Section \ref{subsubsec:303}). This method also bundles all knowledge about ``proper'' dereferencability with redirects in a single place. Another significant advantage of a redirect service is the option to on-board legacy applications fairly easily. In case they already offer URLs that include the accession identifier the time and effort to create permanent URLs is extremely low with a central redirect service. This can slowly move the enterprise towards having permanent URIs for identifying things and, eventually, enables key concepts such as the enterprise knowledge graph and/or data mesh. Finding a good compromise between ``some policies" and ``too many policies" is also crucial in such endeavor. It is clear that a certain level of service availability needs to be guaranteed over a long period of time. In addition, redundancies need to be accounted for as we need to avoid creating a single point of failure.

\begin{infobox}{Software and accession identifiers}
Software is linked to data in the same way as a recipe is linked to its ingredients. Therefore, software---to a large extent---influences how ``resources are identified''. %
In most cases, the role of software is to make these resources accessible for human interaction. In this function of an ``entrance door'' or ``enabler'' (or simply UX) software has received a disproportionate amount of attention; much more than the data itself. It is this imbalance that ``The Data-Centric Manifesto''\footnote{\url{https://web.archive.org/web/20240508005508/http://datacentricmanifesto.org/}} tries to address. Unfortunately, we are not there yet and, as software is close to the user, it typically drives the creation/update/delete cycle for data. This becomes a problem when the process of identifier generation and management become strongly coupled with the software itself. Following are a couple of questions that should play a role when software is created or vendor solutions are chosen:

\begin{enumerate}
   \item Can the new system work with a central identifier service
   \begin{enumerate}
       \item Are accession identifiers part of the an item's URL, e.g., as a GET request parameter?
       \item Does the API offer support to use custom URLs and/or accession identifiers, e.g., \texttt{https://purl.example.com/a9/.*} instead of \texttt{https://sys.example.com/.*} or just the system's own accession identifiers?
   \end{enumerate}
   \item Does the system produce and guarantee persistent accession identifiers? In particular, is there a possibility that previously existing identifiers are re-assigned to new resources after a resource has been deleted?
   \item After system deprecation, can the according accession identifier generation mechanism be reproduced without collision checks (e.g., by increasing a counter or using a prefix)?
\end{enumerate}
\end{infobox}

\subsubsection{Authentication}
We recommend to avoid authentication as much as possible. Of course, in every organization there are areas where (parts of) the handled information is highly sensitive. That, however, does not always mean that the whole application and their URIs need to be locked behind a barrier. Maybe, 99\% of the resource and associated data could actually be freely available within our organization's intranet or even outside. We could then implement authentication and authorization for the remaining 1\% and make the vast majority of our resources available with dereferenceable URIs to ensure easy access for humans and machines alike. Authentication in itself is a dispersed field and, frequently, many mechanisms or interfaces are used within larger organizations. This makes it hard to develop engines that collect data from multiple sources: machines do not only need information about the URIs and associated ontologies but, on top, also need handle the respective authentication barriers. Finally, if we put authentication in place we need to ensure that, during dereferencing, the original request is not lost due to the authentication challenge.

\subsection{Short Stable Paths}
\label{subsec:short-stable-paths}
The path part comes after the top-level domain\footnote{There would still be port numbers in between but we recommend to default to port  443; that's the default for the \texttt{https} scheme and is therefore omitted.} and it includes a series of folder names separated via the \texttt{/} (slash) character and, finally, the accession identifier. It is important to note that the path of any URI is case-sensitive while the scheme, subdomain, and domain parts are not. So 
\begin{verbatim}
https://purl.example.com/a9/e42
\end{verbatim} is considered to identify the same resource
 as \begin{verbatim}
HttPS://PURL.ExAmPlE.com/a9/e42
\end{verbatim}
 but NOT 
 \begin{verbatim}
https://purl.example.com/A9/e42
 \end{verbatim}
We kindly refer the reader to Section 6 of RFC 3986 \cite{URI} and Section 4.2.3 of RFC 9110 \cite{HTTPSemantics} for more information. However, as we are writing this document in relation to ``URIs for FAIR Knowledge Graphs'' it can be safely assumed that RDF is somewhere within the scope of the interested reader. Here we need to {pay attention}: According to Section 3.2 of ``RDF 1.1 Concepts and Abstract Syntax'' \cite{rdf} the recommendation for comparing IRIs (a superset of URIs) in an RDF context boils down to ``Simple String Comaprison'' as defined in Section~6 of RFC 3986 \cite{URI} and Section 5.1 of RFC 3987 \cite{IRI}. This means that we need to be extra careful with case-sensitivity and URL-encoded characters to make our URLs compliant for their use with RDF, not only in the path part, but in all parts of a URL.

Our recommendations for the path include:
\begin{itemize}
\item Make the paths short.
\item Define opaque paths.
\item Stick to lower case.
\item Use only unreserved ASCII characters (as defined in RFC 3986 \cite{URI})
\begin{verbatim}
    unreserved    = ALPHA / DIGIT / "-" / "." / "_" / "~"
\end{verbatim}
\item Use the hyphen character (\texttt{-}) as a default separator (if needed).
\item Avoid ambiguous characters, such as \texttt{[0, O, 1, I, l]}.
\item Avoid trailing endings that point to the technology used such as \texttt{.php}.
\item Avoid trailing slashes.
\end{itemize}
Often there is the desire to communicate some sort of meaning (often in the sense of a ``resource type'') via the folder/context path before finishing with the opaque accession identifier; such as in \url{https://purl.example.com/road/M8} (see \cite{ukgov}). This aspect needs to be considered from multiple angles and there is no right or wrong answer. Here is what we should keep in mind:
\begin{itemize}
    \item Avoid reflecting the (full) type hierarchy in the path: keep it to a {single aspect that never changes} for any individual with an accession identifier that marks the end of that path.\footnote{\label{ft:polarbear}See also: ``Beyond the Polar Bear'' by Mike Atherton (\url{https://youtu.be/8iaqcf9-riI?t=1530})}
    \item Be conscious about competing interests for specific names within your organization. For example, in a bank the word ``account'' is heavily overloaded and there may be multiple claims for exactly this word from different parts of an organization.
    \item Use singular rather than plural.
    \item Consider that you are introducing bias by using one specific language.
\end{itemize}
Generally, if you can, we recommend to stick with opaqueness (so none of these four points apply) and shortness. Lastly, it should be clear that folders are, in fact, optional - we could use accession identifiers right after the top-level domain as is done with \url{https://w3id.org/303}. It is important to note that accession identifiers are part of the path so, although we will cover them in more detail in Section \ref{subsec:accessionidentifiers}, all of the above recommendations also apply to them.

\begin{infobox}{Opaque folder names - erdi8 (Part 1)}
When operating a redirect service (see Section \ref{subsubsec:central_id}) for a whole organization it can be beneficial to avoid mnemonic folder names and assign short name spaces to applications. Opaque folder names maximally decouple the URI scheme from organization and system names. We could start counting with \texttt{a2} in \texttt{erdi8}\footnote{\texttt{erdi8} is another idea that came up during the creation of this manuscript: \url{https://github.com/athalhammer/erdi8-py}} identifier scheme and we get 825 opaque folder options of two-character length only. erdi8 is a reduced Base36 variant that excludes the ``0", ``1'' and ``l" characters to avoid confusion with upper case ``o'' and ``i" respectively (similar to Base58 \cite{Satoshi}):
\texttt{[2 3 4 5 6 7 8 9 a b c d e f g h i j k m n o p q r s t u v w x y z]} In addition, an identifier never starts with a number in that scheme and which makes it compatible to XML QNAMES and also avoids number-only identifiers in that way. 
\end{infobox}

Here are some further resources that touch on the topic:
\begin{itemize}
    \item ``Study on Persistent URIs'' by Phil Archer introduces ``10 DOs and DON'Ts for persistent URIs'' \cite{phil}.
    \item ``Towards a common policy for the governance and management of persistent URIs by EU institutions" recommends two-character namespaces followed by a local identifier (accession identifier).\footnote{\url{https://joinup.ec.europa.eu/collection/semantic-interoperability-community-semic/document/towards-common-approach-management-persistent-http-uris-eu-institutions}}
    \item Google's Knowledge Graph identifiers were constructed with a slash, a single letter prefix (m or g), then slash again, and finally a lowercase sequence of characters and numbers as well as the underscore symbol. In the Google Cloud Knowledge Graph they migrated away from that pattern:
    \begin{quote}
    ``Google Cloud Knowledge Graph introduces a new format of MID, c-024dcv3mk. A single-character namespace c is followed by a hyphen and the base-32 representation of an auto-generated alphanumeric ID. This new form of MID without / provides an easier way for modern applications to integrate."\footnote{\url{https://web.archive.org/web/20221231094434/https://cloud.google.com/enterprise-knowledge-graph/docs/mid}}
\end{quote}
\end{itemize}

\subsection{Opaque Accession Identifiers}
\label{subsec:accessionidentifiers}

Accession identifiers mark the last part of the path and, at the same time, the last part of the URI. It should be clear that all properties that apply to the path in general (see Section \ref{subsec:short-stable-paths}) also should apply to accession identifiers. In other works, accession identifiers are often also called ``local identifier'' or ``local id'' (see \cite{McMurry117812}). We like to think of them in the following categories:

\begin{enumerate}
    \item Mnemonic
    \item Opaque
    \begin{enumerate}
        \item Stateful
        \item Stateless
    \end{enumerate}
\end{enumerate}

\subsubsection{Mnemonic}
In Section 4.5 of ``Cool URIs for the Semantic Web'' Sauermann et al.  recommend to use ``Short, mnemonic URIs...'' \cite{Sauermann}. This has led to a lot of confusion and there is a valid point that usability is improved if a human can quickly grasp ``what is behind the URI''. That said, it also has led to all sorts of abuse. For example, instead of resolving the URI and getting the \texttt{rdfs:label} from the returned RDF data many  developers resorted to simply extract the last part of the URI \texttt{http://dbpedia.org/resource/Douglas\_Adams} and replace the underscore with a space. Of course, this does not necessarily correspond to the \texttt{rdfs:label} but it directly gives the resource a seemingly better name than \texttt{e42}. Unfortunately, this does not consider potential change in the way things are referred to. One example for such name change was the renaming of the ``Republic of Macedonia'' to ``Republic of North Macedonia'' in  2019. The question is what the name change means for the URI  \texttt{http://dbpedia.org/resource/Republic\_of\_Macedonia}.\footnote{This URI exists as of 2021-01-01 and describes the country.} There are two options: 
\begin{enumerate}
    \item Create a new URI that includes the new name and do a \texttt{301 Moved Permanently} redirect from the previous one (see Section \ref{subsubsec:decommission} for more information). Most triples that were associated to the previous URI (e.g., \texttt{rdf:type}, \texttt{dbo:areaTotal}, or \texttt{dbo:capital}) will be transformed to the new URI.\footnote{This is what was actually implemented in January/February 2021.}
    \item Keep the existing URI and just change the \texttt{rdfs:label}.
\end{enumerate}
Both solutions are problematic in their own way:
\begin{enumerate}
    \item To create a new URI is not optimal because the original identifier does not remain stable. This breaks a promise made to external users and also introduces significant overload from a knowledge management perspective (i.e., keeping track of all previously used URIs for each resource; does the order matter?). Metadata becomes tricky to handle: When was the ``new'' document \texttt{http://dbpedia.org/data/North\_Macedonia.ttl} really created? What message does either option transport to our customers? When considering snapshot versioning, there are many triples added and deleted just because the URI of a resource has changed. Incremental/patch versioning is also hard to handle, e.g., do we keep the original changelog when the URI changes?
    \item If the identifier remains stable it will continue to transport incorrect information to human users. For example, in 2020 the HUGO Gene Nomenclature Committee decided to rename the CASC4 (cancer susceptibility candidate 4) gene to GOLM2 in order to move away from the phenotype-related name \cite{gene}. In such a case, where the name changed because science has moved on, it would be highly desirable to have significant uptake from the community. This could be hampered if the  previous name still persists with an incorrect or even misleading accession identifier in the URI.
\end{enumerate}
Mnemonic accession identifiers also make the creators to choose one specific language over others which naturally introduces some sort of bias. There is also the scenario where existing terms get an additional, potentially more prevalent, meaning over time; such as the previously existing terms ``Meta'' and ``Tesla'': both changed their prevalent meaning in the last decades. In such a case it is hard to get around disambiguation resources that handle the new meaning.

A conclusion from this paragraph is that, when choosing mnemonic accession identifiers, there is always a trade-off to be made.
\subsubsection{Opaque}
\label{subsubsec:opaque}
All considerations regarding mnemonic accession identifiers can be avoided by making them opaque from the beginning.
 When a name changes, just the \texttt{rdfs:label} and some other properties may change---but not the URI. When a new, more prevalent, meaning comes up for a term nobody needs to worry about the implications/interpretations for an existing URI with the term in it---it simply does not exist. Opaque accession identifiers are more neutral regarding language\footnote{We are aware that full neutrality can not be achieved as many languages also use their own script. Opting for Latin script and Arabic numerals, as in our recommendation (see Section \ref{subsec:short-stable-paths}), could be considered a discriminating factor.} and do not transport any sort of meaning themselves. 

 Within the category of opaque accession identifiers we can distinguish two categories: stateful and stateless.
\begin{infobox}{URIs in ontologies}
``Should the URIs of ontologies and/or semantic schemas/vocabularies be fully opaque?'' To some extent this is likely the ``Gretchenfrage" of this work. There are two  popular ontologies that use one or the other approach: \texttt{schema.org} uses mnemonic accession identifiers and Wikidata uses fully opaque URIs for instances, classes, and properties. Both works have found significant uptake in the real world. Wikidata has a centralized approach towards using the knowledge graph (edits and queries are done on \texttt{wikidata.org}) while \texttt{schema.org} usage is distributed across the Web. From these two scenarios and their individual focus we can derive the following guideline:
\begin{quote}
Mnemonic URIs can be a better choice for ontologies and/or semantic schemas/vocabularies if tooling for authoring and querying can not be provided in a user-friendly way. 
\end{quote}
However, this also leads to the question whether the current state of a strong URI-focus for common tooling, such as SPARQL UIs, is not a direct result of the long tradition of using mnemonic URIs for ontologies. The reader should therefore reflect on whether they are not adding another stepping stone to this ``vicious cycle'' before choosing mnemonic URIs for their ontology and/or semantic schema/vocabulary.\footnote{See also Mike Pools note on mnemonic accession identifiers in quoted at the end of Section \ref{subsec:accessionidentifiers}.} Part of the path has already been beaten by Wikidata with their visionary approach towards full opaqueness for everything and everyone (stated by \texttt{wikidata:Q51283137}).

\end{infobox}
\paragraph{Stateful}
The use of stateful opaque identifiers is a common pattern and its most simplistic realization is counting. What distinguishes them from stateless identifiers is the fact that uniqueness depends on previously generated identifiers. For example, if you are counting and you want to make sure that you generate a new, unique identifier you need to know the value that was assigned to the previous item; that is the current maximum value which now needs to be increased. Note that, in this paragraph, we do not consider the exact method with which collisions are avoided. For example, you could generate a random identifier but then you need to be either very certain that collisions will not occur (which makes it stateless, see below) or you need to verify that the generated identifier will be unique in the database; for example by scanning its index or probing an insert if there is a unique constraint. From a high-level view it does not matter whether we maintain a \texttt{current\_max} variable for the counter in a memory location or scan a database  index but certainly the former is computationally more efficient.

For stateful accession identifiers we have the following recommendations:

\begin{enumerate}
    \item Keep a counter for resources. Similar to Wikidata we would add a small prefix so it is not just \texttt{42} but \texttt{e42}%
    . However, there are various works that recommend against the use of simple counters \cite{phil}. This is for the fact that you can infer predecessors, successors, and the sequence of creation in general. This can result in undesired effects. For example, at ISWC 2017, Peter Haase pointed out to the author that the sequence of Wikidata's Q-ids can serve as a heuristic ranking of Wikidata items: we just need to consider the numeric part of the Q-ids as a rank---the lower the number the better/higher the rank. The intuition behind this is the idea that items of higher importance are created earlier. That's just another example where Hyrum's Law\cref{hyrum} evidently strikes again. Let us keep that in mind when we consider counting as a viable scheme. As a matter of fact, when counting, the state is defined by the previous identifier (a single variable).
    \begin{enumerate}
        \item Counting can also involve advanced techniques such as counting according to a different base (e.g., Base 32) and also can be limited to a modulo space where then you can start working with a stride parameter. Both techniques can also be combined as implemented in erdi8's (see Infobox on ``Opaque folder names" above) \textit{fancy counting} option.
        \item Leading zeros such as \texttt{e0023} should be avoided when counting in the decimal system. It artificially limits the space\footnote{You can still increase to \texttt{e10000} after \texttt{e9999} but your identifier scheme could be inconsistent eventually---or why would start dropping leading zeros beyond four digits but not 10?} without gain as you probably would not have both, \texttt{e0023} and \texttt{e23}.
    \end{enumerate}
    \item Another way to introduce state is making the complete identifier random or introducing random parts (such as the last three numbers). Here, the difference to \textit{stateless} (see below) is that collision probability is not low enough to confidently use the identifier without checking in the database whether it already exists or has existed before (see Section~\ref{subsubsec:decommission} to learn about decommissioning). In this case, the state is defined by all identifiers that have previously been generated. It is clear that the ``exists or has existed"-check can be done efficiently on a larger number of identifiers only if an according database index is set. We also need to account for a certain number of retries in case we accidentally hit one or more already existing identifiers. The Stanford University Library uses this mechanism\footnote{\url{https://web.archive.org/web/20230916200139/https://raw.githubusercontent.com/sul-dlss/suri-rails/main/app/models/identifier.rb}} for ensuring uniqueness within the set of existing DRUIDs\footnote{\url{https://web.archive.org/web/20230425211238/https://library.stanford.edu/research/stanford-digital-repository/documentation/purls-and-dois}}. As such, the state for random identifiers is the full ledger of identifiers that previously have been produced.
\end{enumerate}
    \begin{infobox}{Opaque accession identifiers - erdi8 (Part 2)}
In the previous information box on ``opaque folder names" we have already introduced \texttt{erdi8}. It is a counting scheme with Base33 and produces predictable identifiers. However, there are advanced features that make the identifiers' appearance more ``random-like'' (where, for example, \texttt{k7zydqrp64} follows \texttt{fmzz7cwc43}) or to re-encode existing identifiers to benefit from features such as the profanity filter (that avoids \texttt{[a, e, i, o, u]}) or shortening existing identifiers that often use a limited alphabet (such as the hex digest of hash functions or UUIDs). 
\end{infobox}
\paragraph{Stateless}
Stateless identifiers bring enough randomness to be able to assume safely that they are not conflicting with (previously) existing identifiers. With stateless identifiers we do not need to consolidate the ledger of existing identifiers to guarantee uniqueness. They can therefore be used directly from the point of generation and can also be produced in a distributed uncoordinated way. As the collision probability depends on the rate of identifier generation (e.g., 1000 per minute) and the total size of the larger search space the boundaries to random stateful identifiers are not clearly set (see above): If we generate a single identifier every 10 years and we have a search space of 100 000 it will be fairly unlikely to generate a collision in the following 100 years. This changes drastically if we would keep the same search space but generate 100 identifiers per second. As such, implementing ``statelessness'' starts with the idea that we trust on the random properties instead of checking on insert.

In addition to pure randomness, stateless identifiers often bring other attributes, such as the number of milliseconds (or microseconds) that have passed from a certain point in time (usually January 01, either 1970 or 1980) in UTC or local time, MAC addresses, hostnames, process identifiers, counters, and namespaces. Even though these elements are often included they are typically obfuscated by some type of encoding so we can still consider them ``opaque". Note that some of these identifier generation mechanisms (e.g., \texttt{xid}s, see below) deliberately introduce an order. When they are sorted proximity means close creation timestamps. 

If there is a dependent object (such as a file or a combination of generally persistent attributes) to identify we can use a variety of different hash algorithms, such as SHA256, MD5, Fowler–Noll–Vo, and others. Note that this is---by definition---not independent of the thing that is being identified. We can assume that this is safe as long as ``whatever is being hashed'' can also be guaranteed to remain unique, there is no further dependency of the type ``the identifier needs to be updated when the thing that was initially hashed has changed (was updated)'', and the collision probability is chosen in accordance to the number of things to be identified. Combining a number of different persistent attributes will provide a good amount of statelessness and opaqueness when the combination is hashed. This is very similar to identifying a good natural key in databases.\footnote{\url{https://en.wikipedia.org/wiki/Natural\_key}}One example for such combination could be ``birth date-time, birth place, and birth name" in the case of a person. More generally, common implementations of stateless identifiers include:

\begin{itemize}
    \item  Universally Unique Identifiers (UUIDs) are 128 bit long and typically encoded as a string that uses hexadecimal characters where blocks are separated by the a hyphen. The version is denoted at the beginning of the third block of characters:\\\texttt{5b6c73a2-b2bc-}\textbf{4}\texttt{11c-c229-9f9gh35ccm1}.\\\ In total there are five different UUID versions. More information on UUIDs and the individual versions can be found in RFC 4122 \cite{UUID}.
    \item \texttt{shortuuid}\footnote{\url{https://github.com/skorokithakis/shortuuid}} is a mechanism that re-encodes UUIDs, that are hex-encoded, to full base64 encoded short versions. This keeps the full information but turns it into significantly shorter strings. Note that your accession identifier will also become case-sensitive. An alternative could be to use \texttt{erdi8} for re-encoding (see info box below on ``Opaque accession identifiers - erdi8'').
    \item \texttt{xid}s\footnote{\url{https://github.com/rs/xid/}} are a combination of seconds since the Unix epoch, machine id (part of hostname hash), process id, and a process-internal counter. They are closely related to MongoDB identifiers and Twitter's Snowflakes\footnote{\url{https://web.archive.org/web/20231215205557/https://blog.twitter.com/engineering/en_us/a/2010/announcing-snowflake}}.  
    \item (Non-)cryptographic hash functions provide uni-directional ways for obfuscating single or combined mnemonic properties representing a natural key. Generally, shorter digests are more user-friendly but at the cost of higher collision probabilities. Making a good choice depends on the identifier creation rate and the general number of resources to be identified.
    \end{itemize}
As the nature of statelessness includes randomness and/or features that set them apart to guarantee uniqueness they typically result in a considerable length (in terms of number of characters). As the general path length should be limited we consider short identifiers, such \texttt{shortuuid}s, erdi8-encoded UUIDs, or \texttt{xid}s as good practice for use cases where statelessness is a key requirement. Hash functions provide the option to make the identifier dependent on one or a combination of properties of the object that is being identified (or even the object itself, in the case of a file). Their primary purpose is to keep uniqueness (limit collisions), obfuscate, and shorten while keeping statelessness. Also a combination of, for example \texttt{xid}s and hash functions is in the range of possibilities. In essence, the key requirement to achieve ``statelessness" is: there is no dependency on the identifiers of any other (potentially) preexisting resources to guarantee uniqueness.

\begin{itemize}
      \item Consult ``Lesson 5. Design new identifiers for diverse uses by others" of~\cite{McMurry117812} for additional pitfalls and guidance on accession identifiers (called ``local identifiers/local IDs'' in that work).
\item Crossref publishes similar instructions on opaqueness and non-predictability as the one presented here for DOI suffixes on their website.\footnote{\url{https://web.archive.org/web/20240324175024/https://www.crossref.org/documentation/member-setup/constructing-your-dois/}}  
\item In Section \ref{subsec:dereferenceability} we introduced the concept of content negotiation and the distinction between identified resources and the documents describing them. We can combine an opaque identifier with a mnemonic slug\footnote{\url{https://en.wikipedia.org/wiki/Clean_URL\#Slug}} for the  URL of the human-readable document that describes the identified resource. In media portals this is often done by adding the name or title of an item as a completely irrelevant part of the URL behind the actual accession identifier. This is also suggested by Mike Atherton: ``301 redirects can be used for a more marketing-friendly URL.''\cref{ft:polarbear}

\item On the ``shortuuid'' repository\footnote{\url{https://github.com/skorokithakis/shortuuid/blob/0b48734be5fe240e5a327ddef9670d86f62a210d/README.md}} the creators---unfortunately---propose: ``If the default 22 digits are too long for you, you can get shorter IDs by just truncating the string to the desired length. The IDs won't be universally unique any longer, but the probability of a collision will still be very low.'' We recommend to ignore this advice as it does not apply generally: depending what version of UUID you use and at what rate you generate identifiers truncating encoded UUIDs on a best-guess base has the potential to set you up for disaster.

\item Michael Pool on mnemonic accession identifiers: ``(...) I think that the usage of meaningful IRIs is a key contributor to what I call the `ontologist's delusion', i.e., the assumption that one has modeled something meaningful because they've created a well-annotated concept that humans can understand. It shouldn't happen, but ontologists really do seem remarkably capable of convincing themselves that they've created meaningful models because they've articulated a clear name in the IRI that the machine processes. An opaque IRI, on the other hand, reminds the ontologist that the only thing the machine has to `understand' the concept is the rules and relations that have been created about that concept.''\footnote{\url{https://web.archive.org/web/20240315214100/https://www.linkedin.com/posts/mikepool_semanticweb-ontology-iri-activity-6939978832795889664-srQG/}}
\end{itemize}

\section{Closing Remarks}
While this work is initially presented as a set of ``do x; avoid y'' statements followed by according rationales it is important to highlight that none of the discussed aspects stand solely for themselves.  For example, a properly administered and set-up redirect service can quickly help over the ``https vs http'' question; URLs of human-readable documents describing a real-world resource can make use of slugs without compromising the opaqueness of the URL that describes the real-world resource; delineating meaning between new and existing URIs can be a complex matter and often there is no straight-forward link: this does not only happen when we want to ``anchor'' our newly coined URI with an existing externally-defined meaning but also when a resource in our own namespace is deprecated and does not have a straight-forward successor. These are examples of why the ``do x; avoid y'' statements in combination quickly lead to a ``whole is greater than the sum of its individual parts'' situation.

There are services and tools that support us while navigating this jungle. For example, the Python library \texttt{rdflib} binds the vocabulary of the most common ontologies so nobody really needs to wonder if the prefix was \texttt{http://schema.org/} versus \texttt{https://schema.org/}; or if the property was \texttt{schema:diseasePreventionInfo} or \texttt{schema:diseasePreventionInformation}; the IDE and code completion abstract from these questions. \texttt{https://w3id.org}  offers a great publicly available redirect service that enables decoupling identifiers from software applications. Many hosting platforms, such as GitHub Pages often deliver the correct media type by analyzing content and/or the file extension. These are great tools that we can directly use or take inspiration from. That said, we are also lacking tooling around some critical parts such as SPARQL UIs that make opaque URIs first-class citizens.

Finally, although this guide provides ``do x; avoid y'' recommendations, creating a good identifier scheme for a use case is not a binary matter. This manuscript should also help the reader to choose the right trade-offs when they are designing URIs as permanent identifiers.

\paragraph{Acknowledgements} The research and manuscript writing was entirely performed in my free time---my job projects in the past four years were just never the right fit for this. However, some of the work I did at Roche had an influence on the thinking and ideas behind the presented concepts. I would like to thank my colleagues Laurent-Philippe Albou, Selena Baset, Javier David Fern\'andez Garc\'ia, Andreas Frank, Sebastian Fr\"ohler, Anna-Kristin Kaufmann, Philip Krauss, Nelia Lasierra Beamonte, Thomas Liener, Paul Murdock, Michal Pietrusinski, Martin Romacker, Joachim Rupp, Felix Schwagereit, Sebastian Scharf, Marija Tiurina, Thomas Weingartner, Martin Zablocki, and Vilius Zabulionis for the good exchange we had around some of the above topics. Further, I would like to thank Mathias Richter and Miel Vander Sade for the good feedback. Thanks also to Bob DuCharme for pointing me to Mike Atherton's presentation ``Beyond the Polar Bear''.
\bibliographystyle{plainnat}
\bibliography{bib}

\begin{thebibliography}{23}
\providecommand{\natexlab}[1]{#1}
\providecommand{\url}[1]{\texttt{#1}}
\expandafter\ifx\csname urlstyle\endcsname\relax
  \providecommand{\doi}[1]{doi: #1}\else
  \providecommand{\doi}{doi: \begingroup \urlstyle{rm}\Url}\fi

\bibitem[Archer(2013)]{phil}
Phil Archer.
\newblock {Study on Persistent URIs}, 2013.
\newblock URL \url{https://philarcher.org/diary/2013/uripersistence/}.

\bibitem[Berners-Lee(1998)]{Berners-Lee1998}
Tim Berners-Lee.
\newblock {Cool URIs don't change}, 1998.
\newblock URL \url{http://www.w3.org/Provider/Style/URI}.

\bibitem[Berners-Lee et~al.(2005)Berners-Lee, Fielding, and Masinter]{URI}
Tim Berners-Lee, Roy~T. Fielding, and Larry Masinter.
\newblock {Uniform Resource Identifier (URI): Generic Syntax}, 2005.
\newblock URL \url{https://tools.ietf.org/html/rfc3986}.

\bibitem[Birbeck and McCarron(2010)]{curie}
Mark Birbeck and Shane McCarron.
\newblock {CURIE Syntax 1.0 - A syntax for expressing Compact URIs}, 2010.
\newblock URL \url{https://www.w3.org/TR/2010/NOTE-curie-20101216/}.

\bibitem[Bruford et~al.(2020)Bruford, Braschi, Denny, Jones, Seal, and
  Tweedie]{gene}
Elspeth~A. Bruford, Bryony Braschi, Paul Denny, Tamsin E.~M. Jones, Ruth~L.
  Seal, and Susan Tweedie.
\newblock Guidelines for human gene nomenclature.
\newblock \emph{Nature Genetics}, 52:\penalty0 754--758, 2020.
\newblock \doi{10.1038/s41588-020-0669-3}.

\bibitem[Clark et~al.(2004)Clark, Martin, and Liefeld]{lsid}
Tim Clark, Sean Martin, and Ted Liefeld.
\newblock {Globally distributed object identification for biological
  knowledgebases}.
\newblock \emph{Briefings in Bioinformatics}, 5\penalty0 (1):\penalty0 59--70,
  03 2004.
\newblock ISSN 1467-5463.
\newblock \doi{10.1093/bib/5.1.59}.

\bibitem[Dürst and Suignard(2005)]{IRI}
Martin Dürst and Michel Suignard.
\newblock {Internationalized Resource Identifiers (IRIs)}, 2005.
\newblock URL \url{https://tools.ietf.org/html/rfc3987}.

\bibitem[Evans(2004)]{ddd}
Eric Evans.
\newblock \emph{Domain-Driven Design: Tackling Complexity in the Heart of
  Software}.
\newblock Addison-Wesley, 2004.

\bibitem[Fielding et~al.(2022)Fielding, Nottingham, and Reschke]{HTTPSemantics}
Roy~T. Fielding, Mark Nottingham, and Julian~F. Reschke.
\newblock {HTTP Semantics}, 2022.
\newblock URL \url{https://tools.ietf.org/html/rfc9110}.

\bibitem[Forkel and Hammarstr\"om(2022)]{glotto}
Robert Forkel and Harald Hammarstr\"om.
\newblock {Glottocodes: Identifiers Linking Families, Languages and Dialects to
  Comprehensive Reference Information}.
\newblock \emph{Semantic Web}, 13\penalty0 (6):\penalty0 917--924, 09 2022.
\newblock \doi{10.3233/SW-212843}.

\bibitem[Halpin(2017)]{Halpin:PrivOn2017}
Harry Halpin.
\newblock Semantic insecurity: Security and the semantic web.
\newblock In Christopher Brewster, Michelle Cheatham, Mathieu d'Aquin, Stefan
  Decker, and Sabrina Kirrane, editors, \emph{5th Workshop on Society, Privacy
  and the Semantic Web - Policy and Technology (PrivOn2017) (PrivOn)}, number
  1951 in CEUR Workshop Proceedings, Aachen, 2017.
\newblock URL \url{http://ceur-ws.org/Vol-1951/#paper-08}.

\bibitem[Halpin et~al.(2010)Halpin, Hayes, McCusker, McGuinness, and
  Thompson]{halpin_identity}
Harry Halpin, Patrick~J. Hayes, James~P. McCusker, Deborah~L. McGuinness, and
  Henry~S. Thompson.
\newblock When owl:sameas isn't the same: An analysis of identity in linked
  data.
\newblock In \emph{The Semantic Web -- ISWC 2010}, pages 305--320, Berlin,
  Heidelberg, 2010. Springer Berlin Heidelberg.
\newblock ISBN 978-3-642-17746-0.
\newblock \doi{10.1007/978-3-642-17746-0\_20}.

\bibitem[Hodges et~al.(2012)Hodges, Jackson, and Barth]{HSTS}
Jeff Hodges, Collin Jackson, and Adam Barth.
\newblock {HTTP Strict Transport Security (HSTS)}, 2012.
\newblock URL \url{https://tools.ietf.org/html/rfc6797}.

\bibitem[Hogan et~al.(2010)Hogan, Harth, Passant, Decker, and
  Polleres]{hogan2010weaving}
Aidan Hogan, Andreas Harth, Alexandre Passant, Stefan Decker, and Axel
  Polleres.
\newblock Weaving the pedantic web.
\newblock In \emph{LDOW}, 2010.

\bibitem[Kahn and Wilensky(1995)]{handle}
Robert Kahn and Robert Wilensky.
\newblock A framework for distributed digital object services, 1995.
\newblock URL \url{http://www.cnri.reston.va.us/k-w.html}.

\bibitem[Leach et~al.(2005)Leach, Mealling, and Salz]{UUID}
Paul Leach, Michael Mealling, and Rich Salz.
\newblock {A Universally Unique IDentifier (UUID) URN Namespace}, 2005.
\newblock URL \url{https://tools.ietf.org/html/rfc4122}.

\bibitem[Lewis(2007)]{httpRange-14}
Rhys Lewis.
\newblock {Dereferencing HTTP URIs (W3C Draft)}, 2007.
\newblock URL
  \url{https://www.w3.org/2001/tag/doc/httpRange-14/2007-08-31/HttpRange-14.html}.

\bibitem[McMurry et~al.(2017)]{McMurry117812}
Julie~A McMurry et~al.
\newblock Identifiers for the 21st century: How to design, provision, and reuse
  persistent identifiers to maximize utility and impact of life science data.
\newblock \emph{PLOS Biology}, 15\penalty0 (6):\penalty0 1--18, 06 2017.
\newblock \doi{10.1371/journal.pbio.2001414}.

\bibitem[Nakamoto and Sporny(2020)]{Satoshi}
Satoshi Nakamoto and Manu Sporny.
\newblock {The Base58 Encoding Scheme}, 2020.
\newblock URL \url{https://tools.ietf.org/id/draft-msporny-base58-02.txt}.

\bibitem[of~the CTO Council’s~cross Government
  Enterprise~Architecture(2009)]{ukgov}
Public Sector Information~Domain of~the CTO Council’s~cross Government
  Enterprise~Architecture.
\newblock {Designing URI Sets for the UK Public Sector}, 2009.
\newblock URL
  \url{https://www.gov.uk/government/publications/designing-uri-sets-for-the-uk-public-sector}.

\bibitem[Recommendation(2014)]{rdf}
W3C Recommendation.
\newblock {RDF 1.1 Concepts and Abstract Syntax}, 2014.
\newblock URL \url{https://www.w3.org/TR/rdf11-concepts}.

\bibitem[Sauermann et~al.(2008)Sauermann, Cyganiak, V{\"o}lkel, and
  Ayers]{Sauermann}
Leo Sauermann, Richard Cyganiak, Max V{\"o}lkel, and Danny Ayers.
\newblock {Cool URIs for the Semantic Web}, 2008.
\newblock URL \url{https://www.w3.org/TR/cooluris/}.

\bibitem[Wilkinson et~al.(2016)Wilkinson, Dumontier, Aalbersberg, et~al.]{fair}
{Mark D.} Wilkinson, Michel Dumontier, {Ijsbrand Jan} Aalbersberg, et~al.
\newblock The fair guiding principles for scientific data management and
  stewardship.
\newblock \emph{Scientific Data}, 3, March 2016.
\newblock ISSN 2052-4463.
\newblock \doi{10.1038/sdata.2016.18}.

\end{thebibliography}

\appendix
\section{Prefixes used in this work}
\label{apx:prefix}
\begin{itemize}
    \item \texttt{dcterms} - {http://purl.org/dc/terms/}
    \item \texttt{dbo} - {http://dbpedia.org/ontology/}
    \item \texttt{ex} - {https://purl.example.com/a9/}
    \item \texttt{owl} - {http://www.w3.org/2002/07/owl\#}
    \item \texttt{schema} - {https://schema.org/}
    \item \texttt{skos} - {http://www.w3.org/2004/02/skos/core\#}
    \item \texttt{uniprot} - {http://purl.uniprot.org/uniprot/}
    \item \texttt{wikidata} - {http://www.wikidata.org/entity/}
    \item \texttt{wdrs} - {https://www.w3.org/2007/05/powder-s}
\end{itemize}

\end{document}